%% file: main.tex
\documentclass[a4paper,11pt]{article}
\pdfoutput=1 

\usepackage{graphicx} 
\usepackage{circuitikz}

\usepackage{xfrac}
\usepackage{xcolor}
\usepackage{soul}
\usepackage{tcolorbox}
\usepackage{makecell}

\usepackage{subcaption}
\usepackage{array}
\newcolumntype{P}[1]{>{\centering\arraybackslash}p{#1}}
\usepackage{floatrow}
\graphicspath{ {./images/} }

\usepackage{jinstpub} 

\usepackage{amsmath}  
\usepackage{lineno}

\usepackage[utf8]{inputenc}
\usepackage{orcidlink}
\usepackage{siunitx}

\DeclareSIUnit\curie{Ci}

\title{Red Pitaya DAQ for fast neutron detection: a scalable, compact and low-cost solution}

\author[a,1]{Dzmitry Kazlou,\orcidlink{0009-0002-2658-2347} \note{Corresponding author.}}
\author[a]{Roman Bergert,\orcidlink{0000-0003-4109-7534}}
\author[a]{Hans-Georg Zaunick,\orcidlink{0000-0002-6835-0109}}
\author[a]{Kai-Thomas Brinkmann,\orcidlink{0000-0002-4387-8193}}

\affiliation[a]{$2^{nd}$ Physics Institute Justus Liebig University, Giessen, Germany}

\emailAdd{dzmitry.kazlou@exp2.physik.uni-giessen.de}

\abstract{In this work, we present the initial results of a compact fast-neutron detection system based on the EJ-276D plastic scintillator, a dual-channel readout using two silicon photomultipliers (SiPMs) with complementary cell geometries, and a data acquisition (DAQ) system built on the StemLab Red Pitaya FPGA platform. The system combines the pulse shape discrimination (PSD) capability of EJ-276D with custom, cost-effective analog front-end electronics and open-source DAQ firmware. Unlike traditional systems relying on bulky photomultiplier tubes or liquid scintillator detectors, our design is portable and scalable. It achieves a Figure of Merit of 1.61 in the 2.75--3.0\,MeV$_{ee}$ energy range, comparable to established setups, while requiring only a single +5\,V USB power source. The results demonstrate the potential of this approach for in-situ neutron detection in mixed radiation fields, with the FPGA programmability of the Red Pitaya enabling straightforward adaptation to diverse academic and industrial applications.}

\begin{document}

\maketitle

\section{Introduction}
\label{Sec:Introduction}

Neutron detection is a cornerstone of numerous scientific and practical applications, ranging from nuclear physics research and homeland security to medical imaging and subsurface logging \cite{ref:solar, ref:imaging, ref:homesec, ref:medapp}. Due to their lack of electric charge, neutrons interact with matter primarily through nuclear processes, necessitating specialised detection techniques. These techniques can be broadly classified into two categories based on neutron energy: thermal and fast neutron detection. Thermal neutron detection is based on materials containing isotopes with high thermal neutron capture cross-sections, such as $^3$He, $^6$Li, $^{10}$B or $^{157}$Gd. Some of these nuclides produce detectable charged particles upon neutron capture. In contrast, fast neutron detection is based on the neutron proton (n,p) interaction, in which energetic neutrons scatter off hydrogen nuclei to generate recoil protons that produce scintillation light in hydrogen-rich materials \cite{ref:knoll}.

A significant challenge in fast neutron detection is the separation of neutron-induced signals from gamma-ray backgrounds, which are present in mixed radiation environments. Pulse Shape Discrimination (PSD) is a critical technique that addresses this issue by exploiting the differences in the shapes of the scintillation signals produced by neutrons and gamma rays \cite{ref:knoll}. Organic scintillators, including liquid scintillators (e.g. EJ-301, EJ-309) and solid plastic scintillators (e.g. EJ-276D, EJ-276G), and certain scintillation glasses, have been extensively studied for their PSD capabilities \cite{ref:recent2,ref:EJ301_PSD,ref:ej276,ref:fast_neutron_psd}. A comparison of standard liquid and solid plastic materials with organic glasses was carried out by Laplace et al.\,\cite{ref:comparative}. Among these, stilbene has historically been regarded as a benchmark material due to its superior PSD performance \cite{ref:stilbenech}. However, traditional detection systems employing stilbene or liquid scintillators often suffer practical limitations, including fragility and the requirement for specialised handling equipment and bulky readout systems that necessitate high-voltage power supplies and expensive DAQ hardware.

Recent advances have focused on developing compact, cost-effective, and robust neutron detection systems that use commercially available materials and SiPM photosensors \cite{ref:ej276_sipm_psd,ref:buffler2023,ref:SiPM_performance}. When paired with modern DAQ platforms such as the StemLab Red Pitaya, SiPM-based readout systems offer a versatile and affordable solution for neutron detection.

The StemLab Red Pitaya 125-14 platform is built on an AMD Xilinx Zynq 7000 series System-on-Chip (SoC) with a dual-core ARM Cortex-A9 processor. It integrates a field-programmable gate array (FPGA) and dual- or quad-input channels with 125~MS/s 14-bit analog-to-digital converters (ADCs), and runs a lightweight Linux operating system~\cite{ref:redpitaya}. This architecture supports high-speed signal processing with a sampling clock resolution of \SI{8}{\nano \second}, direct memory acquisition (DMA) with expandable buffer size and a total dead time of \SI{196.61}{\micro\second}, which allows us to deal with event rates up to \SI{5000}{\per\second}.

The real-time data acquisition makes it ideal for PSD and other advanced signal analysis techniques. Studies using the Red Pitaya for radiation detection have demonstrated its efficiency in processing high-frequency signals with minimum latency \cite{ref:RP1, ref:RP2, ref:RP3, ref:RP4}. The use of commercially available components ensures scalability and accessibility for a wide range of applications. This article details the experimental setup, signal processing methodology, and preliminary results, highlighting the performance and advantages of this approach.

\section{Instrumentation and Methods}
\label{Sec:MaterialsAndMethods}

\begin{figure}[h!]
\centerline{%
\includegraphics[width=0.6\textwidth]{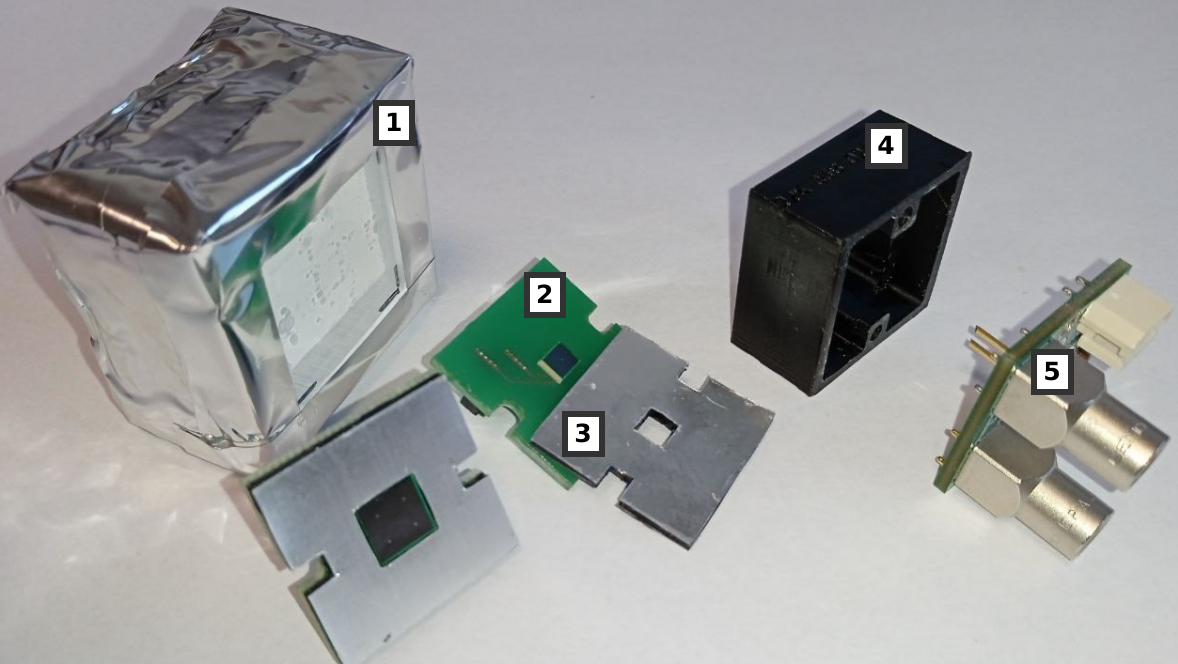}
\includegraphics[width=0.4\textwidth]{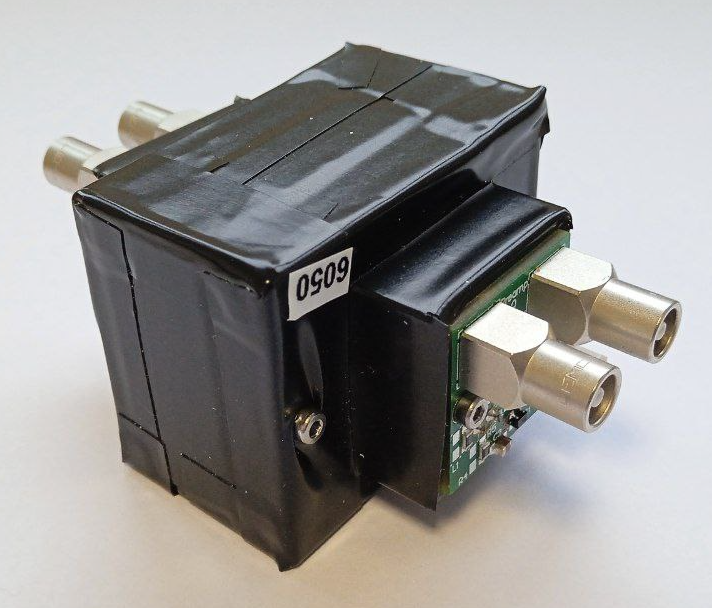}
}
\caption{\label{fig:exploded} 
Detector building blocks consisting of scintillator (1) wrapped in PTFE and reflective foil, SiPM PCB (2) with levelling mask (3), preamplifier (5) with 3D-printed holder (4, left) and fully assembled detector head (right).
}
\end{figure}

\subsection{Detector Material}
\label{Sec:DetectorMeterial}

The commercially available EJ-276D plastic scintillator (Eljen Technology)~\cite{ref:ej276d, ref:pant}  was chosen as the detector material due to its well-performing pulse shape discrimination (PSD) capability, enabling efficient separation of fast neutrons from $\gamma$-rays. A sample with the dimensions of \numproduct{3.2 x 3.2 x 2} cm$^3$ was wrapped with multiple layers of PTFE and one additional outer layer of aluminum foil. 
PTFE is a well-established diffuse reflector for organic scintillators, providing reflectivity typically exceeding 95\% in the relevant wavelength range while being chemically inert and easy to  handle~\cite{ref:teflon_wrapping,ref:teflon_wrapping2,ref:teflon_wrapping3}. Multiple layers are used to ensure uniform reflectivity and to avoid light leakage through thin spots while the outer aluminum foil layer provides additional light tightness.  
Both SiPMs were optically coupled with silicone grease (Bayer Baysilone M300000). All components were assembled into a 3D-printed housing structure and additionally wrapped with duct tape for light tightness. A photograph of all detector parts and the assembled detector head is shown in Figure~\ref{fig:exploded}.

\subsection{Readout Electronics}
\label{Sec:ReadoutElectronics}

The choice of photo sensors and the following front-end electronics was determined by two main requirements: a) a large dynamic range to cover energies up to the full energy deposit of protons along the thickness of the sensitive volume and b) a high sensitivity to allow for weak signal detection and the possibility of energy calibration using gamma emitters. Both requirements are mutually exclusive when utilizing single modern SiPMs as photosensors due to the limited number of cells on one hand or, conversely, because of the low gain when covering large light signals on the other hand.
Therefore, a dual-channel readout of the scintillator by SiPM photosensors with different properties was realized. The first channel comprises a Hamamatsu S14160-6050HS with an active area of \numproduct{6 x 6} \si{\square\mm}, 
\num{14331} cells and a cell pitch of \SI{50}{\micro\meter}~\cite{ref:hamasipm1}. This SiPM type is deemed advantageous for low amplitude signals due to its large sensitive area and superior photon detection efficiency (PDE) of \SI{50}{\percent}. The second channel uses a Hamamatsu S14160-3010PS with a sensitive area of \numproduct{3 x 3} \si{\square\mm}, \num{89984} cells and a cell pitch size of \SI{10}{\micro\meter}~\cite{ref:hamasipm2}. Besides the reduced overall area, this choice promises a larger dynamic range due to the larger amount of cells compared to the \SI{50}{\micro\meter} pitch SiPM.

This dual-channel readout scheme effectively extends the dynamic
range to the physical limit given by the largest expected energy deposit
inside the detector material while preserving the excellent low-energy
response below \SI{1}{\mega\electronvolt} owing to the large area and cell pitch size of the S14160-6050HS SiPM. Concretely, the high-sensitivity channel is designed to cover the range from 0 to 8\,MeV$_{ee}$, appropriate for the recoil proton spectra of standard laboratory neutron sources ($^{252}$Cf, Ra-Be, Am-Be), while the high dynamic range channel is designed to extend this coverage to energies well above \SI{100}{\mega\electronvolt_{ee}}. The practical upper limit of each channel was confirmed experimentally during the $^{12}$C beam measurements described in Section~\ref{Sec:ResultsAndDiscussion}.

\begin{figure}[h]
\centering
\begin{minipage}[t]{0.48\textwidth}
    \centering
    \resizebox{\linewidth}{!}{\input{preamp1}}
\end{minipage}%
\hfill
\begin{minipage}[t]{0.48\textwidth}
    \centering
    \resizebox{\linewidth}{!}{\input{preamp2}}
\end{minipage}
\caption{\label{fig:preamps}
Schematics of SiPM preamplifiers matched to Hamamatsu S14160-6050HS (left) 
and S14160-3010PS (right).
}
\end{figure}
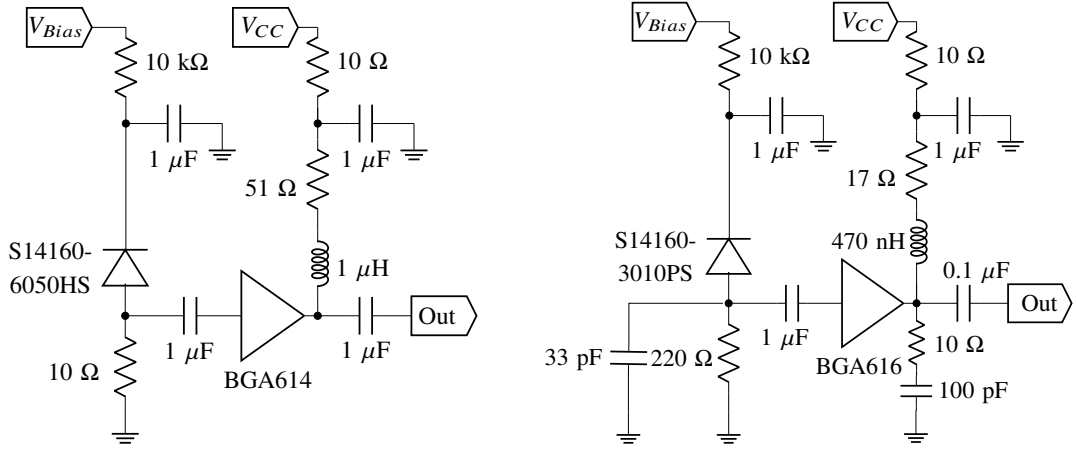

Broadband amplifiers based on the BGA61x family of silicon monolithic microwave integrated circuit (MMIC) amplifiers by Infineon Technologies~\cite{ref:bga614} have been developed and matched to the specific characteristics of the SiPMs used. The circuits were tuned to avoid overshoot while maintaining an acceptable decay time leading to a signal length that matches the sampling resolution of Red Pitaya's onboard ADC. The schematics of both such tuned preamplifiers are shown in Figure~\ref{fig:preamps}.

\begin{figure}[h]
\centerline{
    \includegraphics[width=1.0\textwidth]{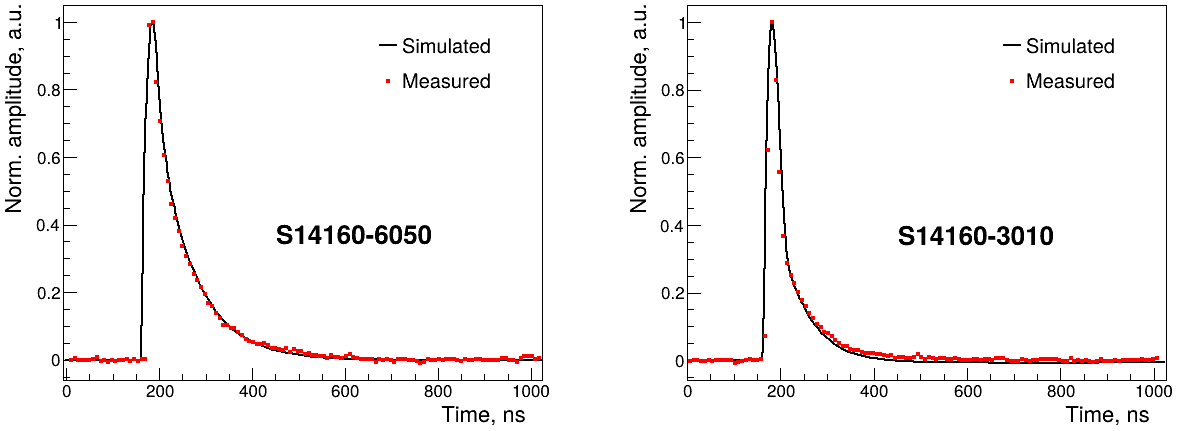}
}
\caption{\label{fig:simvsexp} 
Examples of simulated and experimentally measured raw traces for both channels based on SiPMs Hamamatsu S14160-6050HS (left) and S14160-3010PS (right).
}
\end{figure}

For the first (high sensitivity) channel, the SiPM anode-side resistor was set to a rather small value of \SI{10}{\ohm} to ensure a fast fall-off time and overall reduced amplitude. No additional filtering or pole-zero compensation is necessary owing to the large SiPM capacitance. The signals of the second (high dynamic range) channel, on the other hand, are in comparison smaller and shorter due to the small cell sizes of the S14160-3010PS resulting in a tiny amplifier input capacitance. Here, additional low-pass filtering is achieved by the parallel combination of the anode-side network to increase the signal-to-noise ratio (SNR). A trap filter at the MMIC output dampens potential ringing at frequencies beyond the signal band. For the MMIC, a BGA616 (in contrast to the first channel's BGA614) was chosen in order to provide sufficient swing covering the large dynamic range. It is worth to note that the self-resonant frequency (SRF) of the inductivities used should be greater than the \SI{3}{\decibel}-bandwidth of the MMICs for the circuits to exhibit proper function. The use of RF-rated components and layout techniques is highly recommended.

The preamplifier circuit was simulated by the LTSpice framework \cite{ref:LTSpice} to adapt the pulse shape to the Red Pitaya's ADC sampling resolution. A successful modification to stretch the pulse lengths could be achieved while mitigating overshoots or ringing in general without compromising too much gain. The optimum of the simulated pulses were crosschecked with acquired signals of each channel and showing remarkable overlap as shown in Figure~\ref{fig:simvsexp}.

Aiming to have an independent detector assembly from external high voltage power supplies and to make the entire project compact and self-contained, a printed circuit board (PCB) based on the LT8362 DC/DC converter \cite{ref:lt8362} was designed for the SiPM bias voltage supply. The circuit is a modified version of the bias voltage supply solution of the open source project \textit{OpenSiPM} \cite{ref:biaspcb_github,ref:biaspcb}. With the used bias voltage supply solution and the +5 V from the GPIO pin header of the Red Pitaya board all necessary power is distributed for the preamplifier and +43 V bias for the SiPMs. Hence, the whole detector assembly requires only USB power supplied through the Red Pitaya's USB power port, which can be achieved by either the native power plug or potentially by a power bank.

\subsection{Data Acquisition and Processing}
\label{Sec:DataAcqusition}

The preamplifier outputs are acquired by the Red Pitaya's onboard ADCs with proper \SI{50}{\ohm} impedance matching at the inputs, which lack internal termination resistors. The 14 bit ADC, configured for an input range of \SI{\pm 1}{\volt}, captures amplified signals with 128 samples corresponding to a trace length of \SI{1024}{\ns}.  
All measurements were performed with the basic FPGA configuration, without the original library functionality to avoid all unnecessary calculations that would slow down the data acquisition flow and to deal with raw data. For that purpose a simple but effective C code was written based on Red Pitaya v0.94 FPGA register map. The program placed into a circular buffer of DDR memory only the triggered signals fitting threshold requirements that were chosen to lie above the SiPM dark count noise floor and stored them continuously into a binary file locally on the Red Pitaya's SD card. Then saved data were remotely accessed and copied for further full offline analysis. The latter is performed with the CERN ROOT framework~\cite{ref:root}.

Furthermore, the code has been updated to a full-fledged server service that communicates with the Red Pitaya board via Ethernet by the TCP protocol. The software, in addition to sending configuration commands to the board, allows operation in two modes: retrieving raw traces, or performing additional analysis such as baseline subtraction, extraction of the maximum amplitude, and pulse integral which is performed online and only the output histograms are sent to the user, significantly reducing output data throughput.

The basic FPGA configuration provided by the manufacturer includes a split trigger mode that should allow simultaneous triggering of all channels; however, in practice, only sequential traces from two channels will be captured. This drawback was corrected in a specially designed FPGA firmware, which, along with the entire software package, is available as open source~\cite{ref:drkazlou_github}.

\subsection{Energy Calibration}
\label{Sec:Calibration}

\begin{figure}[htbp]
\centerline{%
\includegraphics[width=1.0\textwidth]{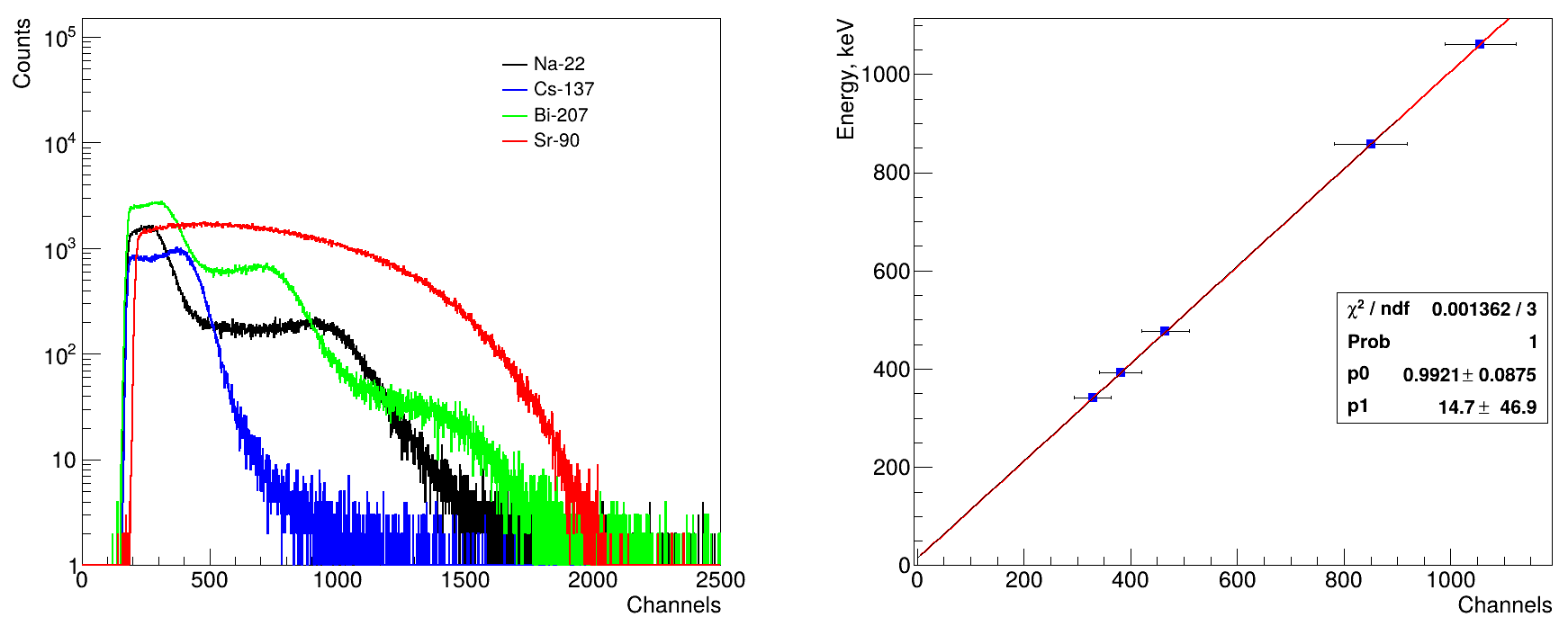}
}
\caption{\label{fig:calib} 
Amplitude spectra obtained using standard gamma sources  and linear calibration of ADC channels into units of MeV electron equivalent ($MeV_{ee}$). The error bars represent $\sigma$ values of Compton edge fitted according to eqn. \ref{eq:calib}.
}
\end{figure}

Due to the limited dimensions and the inherently low effective charge number of the detector, an energy calibration following the traditional approach of fitting the $\gamma$ emission's photo peak as energy reference fails, since the majority of interactions inside the detector is dominated by Compton scattering but negligible photo absorption. Therefore, the energy spectra (Figure~\ref{fig:calib}) are characterized by pure Compton-electron recoil distributions with no discernible photo peak, precluding a conventional peak-based calibration. Thus, the calibration is performed using the Compton edge, yielding a unified electron-equivalent energy scale that is equal for electrons and $\gamma$-quanta.

Hence, we determine the position of the Compton edge from the obtained spectra, with the maximum electron energy $E_{e,\max}$ being related to the emitter’s known gamma energy $E_\gamma$ by
\begin{equation}
 \tag{1}
 \label{eq:compton}
E_{e,\max} = \frac{2 E_\gamma^2}{m_e c^2 + 2 E_\gamma},
\end{equation}
with the electron's rest mass energy $m_e c^2$ = \SI{511}{\kilo\electronvolt}.

The positions of the Compton edges were first estimated by numerically differentiating the amplitude spectra and subsequently finding local minima as proposed in \cite{ref:calibtech,ref:ej276_calib}. The so obtained coarse positions were then used as initial parameters for fitting the following function to the amplitude spectra: 

\begin{equation}
    \tag{2}
    \label{eq:calib}
    f(x) = A \cdot \operatorname{Erfc}\left(\frac{x - x_0}{\sqrt{2}\sigma}\right) + Bx + C
\end{equation}
with the fit parameter $x_0$ yielding the Compton edge position.

\begin{table}[htbp]
    \centering
    \begin{tabular}{ccccc}
         \hline
         Nuclide&$E_\gamma$\,(keV) & $E_{e,\max}$\,(keV) & M (ADC ch.) & $\sigma$ (ADC ch.)  \\
         \hline
         $^{22}$Na & 511 & 341 & 329 & 34.6\\
                   & 1275 & 1061 & 1055 & 66.7\\
         $^{137}$Cs & 662 & 478 & 465 & 44.8\\
         $^{207}$Bi & 569 & 394 & 381 & 39.2\\
                    & 1063 & 860 & 850 & 68.6\\
         \hline
    \end{tabular}
    \caption{Gamma emissions with Compton edge energies used for
electron-equivalent calibration. M and $\sigma$ are the Gaussian
fit parameters extracted from the ADC amplitude spectra; these values
correspond to the calibration points shown in Figure~\ref{fig:calib}.}
    \label{tab:gamma_energies}
\end{table}

In order to obtain an energy calibration for electron equivalent energy deposit, amplitude spectra for several radioactive sources were recorded and the Compton edges fitted with the algorithm described above.

Table~\ref{tab:gamma_energies} gives an overview of the utilized
emitters together with their gamma emission energies, the corresponding
maximum Compton electron energies $E_{e,\max}$, and the Gaussian fit
parameters (M, $\sigma$) extracted from the measured ADC amplitude
spectra. The fitted peak positions M serve as the calibration points
mapping ADC channels to electron-equivalent energy, while $\sigma$
characterizes the energy resolution of the detector at each reference
energy.

The raw amplitude spectra obtained for each of the calibration sources is shown in Figure~\ref{fig:calib} together with the calibration resulting from the Compton edge calibration procedure. In addition, the detector response to the $\beta^-$ emission of a $^{90}$Sr source with a maximum electron energy of \SI{2.28}{\mega\electronvolt} was obtained. The endpoint of the $\beta^-$ energy spectrum (maximum electron kinetic energy of 2.28\,MeV) is used to achieve inter-calibration between the two detector channels, since above calibration based on gamma sources is not applicable to the second (high dynamic range) channel due to its insensitivity in the low energy range.

\subsection{Pulse Shape Discrimination Method}
\label{Sec:PSD}
\begin{figure}[htbp]
\includegraphics[width=0.5\textwidth]{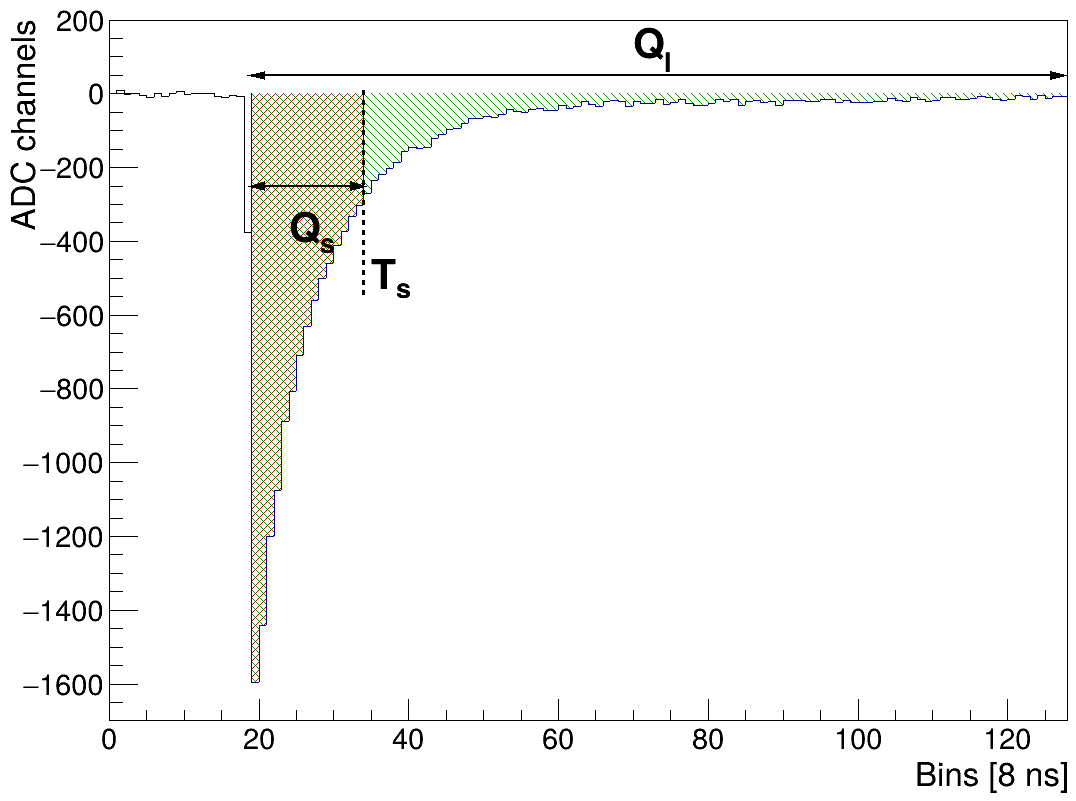}
\includegraphics[width=0.4\textwidth]{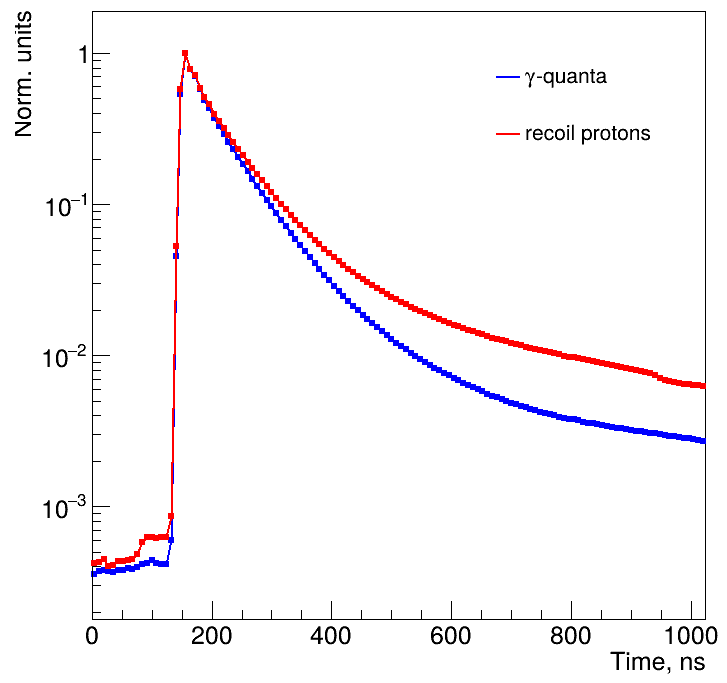}
\caption{\label{fig:psdexplanation} 
Illustration of the PSD parameter definition (left) and experimentally obtained averaged traces (right) for recoil protons and $\gamma$-quanta.}
\end{figure}

During offline analysis, for each raw waveform base line subtraction, amplitude extraction, calculation of partial ($Q_s$) and full ($Q_l$) integrals (see Figure~\ref{fig:psdexplanation}) are performed. 

From latter quantities, the Pulse Shape Discrimination parameter (PSD) as an observable for the desired n/$\gamma$ separation is computed as shown on Figure~\ref{fig:psdexplanation}
\begin{equation}
    \tag{3}
    \label{eq:psd}
    PSD = 1 - \frac{Q_s}{Q_l}.
\end{equation}

\begin{figure}[htbp]
\includegraphics[width=0.5\textwidth]{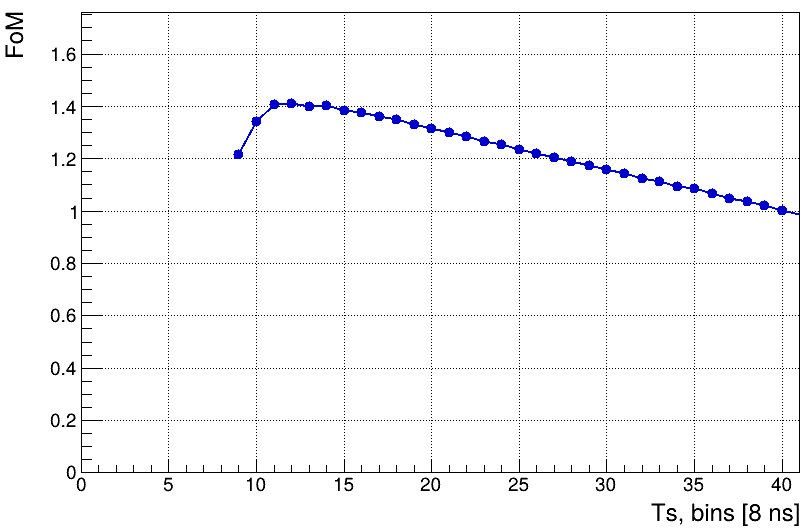}
\caption{\label{fig:fomvsts} 
Illustration of the $T_s$ parameter optimization. The error bars are smaller than the marker size.}
\end{figure}

For each recorded waveform, amplitude and PSD values are extracted and stored in a two dimensional PSD-vs-amplitude histogram. The emerging distributions are characterized through statistical methods.
A Figure of Merit (FoM) for the n/$\gamma$ discrimination at a fixed amplitude based on the extracted PSD value can then be defined by
\begin{equation}
    \tag{4}
    \label{eq:fom}
    FoM = \frac{M_n - M_\gamma}{FWHM_n + FWHM_\gamma},
\end{equation}
where $M_n$ and $M_\gamma$ are the centers of the neutron and gamma Gaussian fits, $FWHM_n$ and $FWHM_\gamma$ are the full width at half maximum (FWHM) values of the neutron and gamma peaks in PSD space, respectively. The center of the peaks and the respective widths were estimated using a Gaussian fit.

A systematic scan of FoM versus $T_s$ was performed on a dedicated subset of the Am-Be dataset in the energy range of \SI{1.5}-\SI{2.0}{\mega\electronvolt}, varying $T_s$ from 9 (72\,ns) to 40 bins (320\,ns) in steps of one bin, as shown in Figure~\ref{fig:fomvsts}. The value $T_s = 12$ bins (\SI{96}{\nano\second}), measured from the position of the maximum amplitude sample, was found to maximise the discrimination of n/$\gamma$ bands {across the energy range of interest} and therefore was fixed for all subsequent measurements.

Data sets with sufficient statistics to allow a particle discrimination and determination of the corresponding FoM were obtained with laboratory neutron sources as described in section \ref{Sec:LabTests}.

\section{Laboratory Tests}
\label{Sec:LabTests}
Initial tests of the pulse shape discrimination capability were conducted using laboratory neutron sources. For these measurements, three different neutron sources were utilized:
\begin{itemize}
    \item A mixed nuclide $^{250}$Cf/$^{252}$Cf source with a current neutron emission rate of \SI{2e5}{\per\second}.
    \item A Ra-Be howitzer with source containing \SI{3}{\mg} $^{226}$Ra and nominal neutron integral flux of \SI{5e4}{\per\second} enclosed by a paraffin mantle. The measurement setup was positioned over the tangential channel closest to the source in order to maximize the fast-to-thermal neutron ratio.
    \item An AmBe source with a nominal unmoderated neutron flux of \SI{1.3e7}{\per\square\cm\per\second} at the default measurement position.
\end{itemize}
The measurement duration was adjusted depending on each source's activity to obtain a comparable number of total events. The time intervals were 65 hours, 15 hours and 10 minutes for the $^{250}$Cf/$^{252}$Cf, Ra-Be and Am-Be neutron sources, respectively. The long duration measurement with the californium source was performed in a temperature stabilized laboratory at 20\,$^\circ$C. No systematic drift in energy calibration, baseline or PSD peak position was observed within any single measurement campaign. Temperature monitoring and active bias compensation are planned for the next hardware revision. 

\section{Results and Discussion}
\label{Sec:ResultsAndDiscussion}
\begin{figure}[htbp]
\centerline{\includegraphics[width=0.5\textwidth]{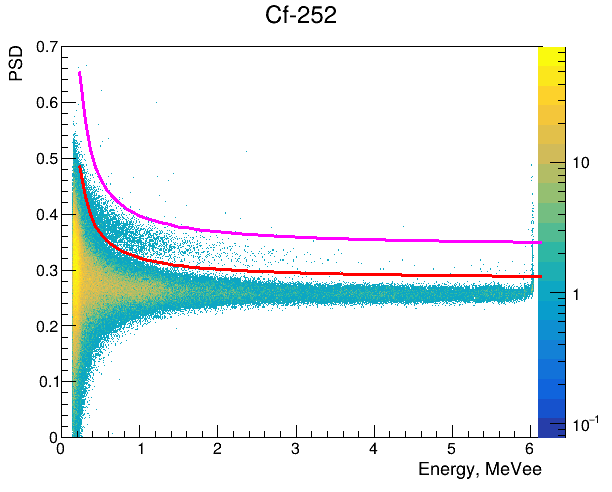}
\includegraphics[width=0.5\textwidth]{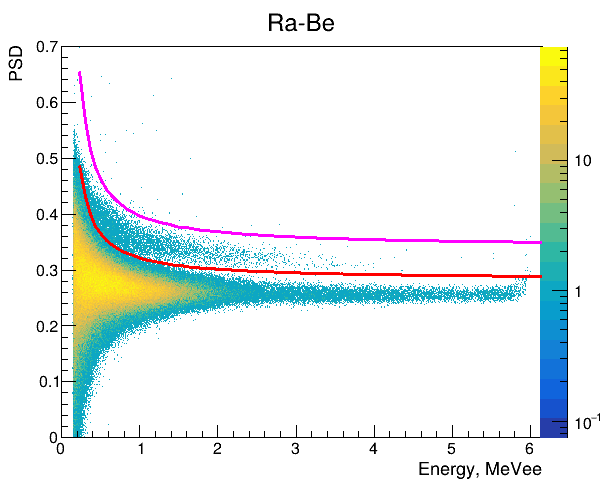}}
\centerline{\includegraphics[width=0.5\textwidth]{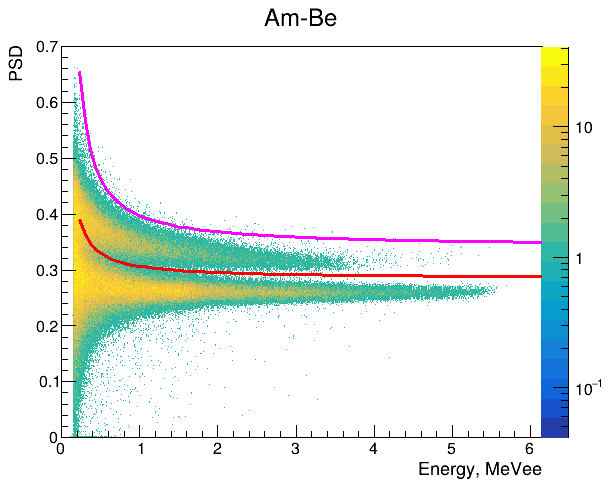}
\includegraphics[width=0.5\textwidth]{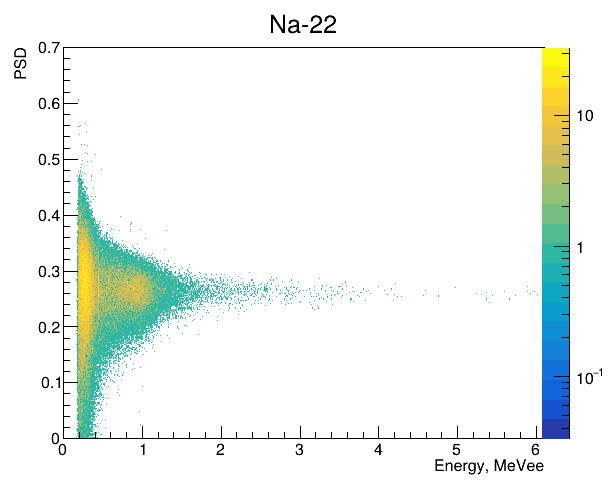}}
\caption{\label{fig:psd2d} 
Scatter-plot of PSD values versus deposited energy in MeV electron equivalent measured with $^{252}$Cf, Ra-Be, Am-Be neutron sources and $^{22}$Na~$\gamma$-source.}
\end{figure}

For each measurement, waveforms were analyzed as described in section~\ref{Sec:PSD}. With the energy calibration (section~\ref{Sec:Calibration}) applied, PSD-vs-energy histograms are obtained as shown in Figure~\ref{fig:psd2d}. Each panel uses an independent color scale, representing local event density of the specific dataset.

\begin{figure}[h]
\includegraphics[width=0.47\textwidth]{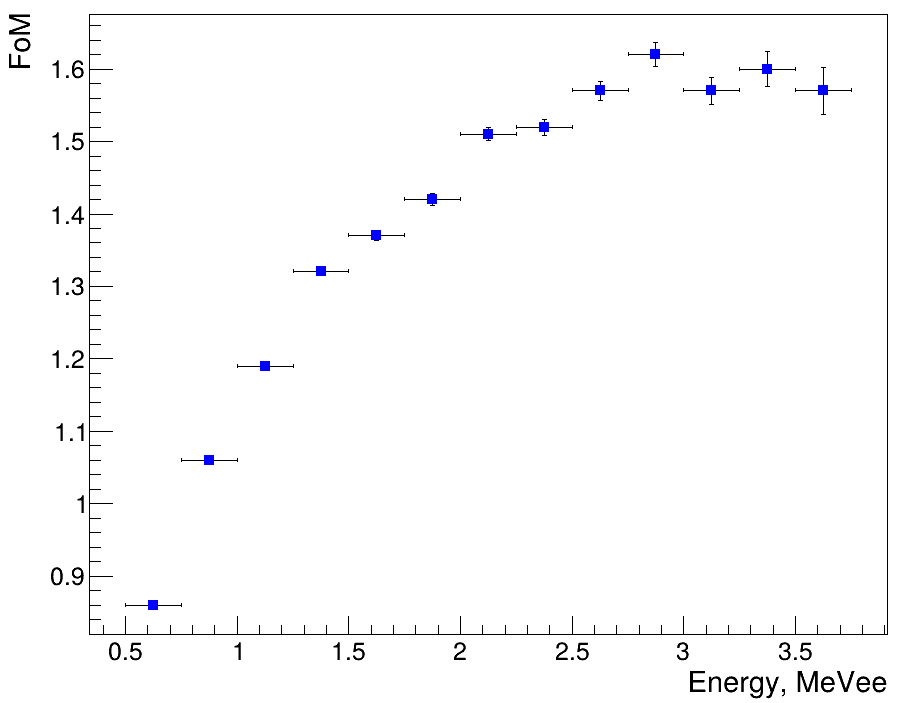}
\includegraphics[width=0.47\textwidth]{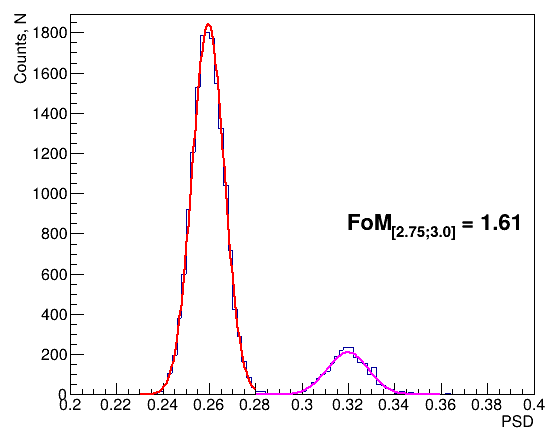}
\caption{\label{fig:psd1d} 
FoM as a function of energy range and example of PSD histograms for 2.75-3.0 $MeV_{ee}$ cut-out with Am-Be neutron source. 
}
\end{figure}

Two distinct areas can be recognized in the plots visually separated by the red curve. The lower band corresponds to events from $\gamma$-quanta that are abundantly produced as a result of radioactive decay in the neutron sources used. The upper distribution is attributed to recoiling protons originating from elastic (n,p) reactions. For further analysis, all events inside the area defined by red and magenta boundary lines are treated as fast neutron events. Finally, both distributions are investigated separately at fixed energy intervals. 
Projections of the distributions in the energy range \SI{0.5}-\SI{3.5}{\mega\electronvolt} with a width of \SI{0.25}{\mega\electronvolt} were fitted with Gaussians, and the FoM parameter was extracted. The dependence of the obtained FoM values on the midpoint energy of each interval, together with an example of a single FoM calculation is presented in Figure~\ref{fig:psd1d}. 

The best n/$\gamma$ separation performance was observed in the \SI{2.75}-\SI{3.0}{\mega\electronvolt} interval, yielding a FoM of 1.61, which compares well with other published values~\cite{ref:buffler,ref:wang, ref:pant}. Averaged pulse shapes, normalised by amplitude, for the two distinguished populations (n,p) reactions and $\gamma$-quanta are shown in Figure \ref{fig:psdexplanation} on the right.

Additional measurements were conducted using a \numproduct{30 x 30 x 41} cm$^3$ water phantom \cite{ref:phantom}. This setup enabled the generation of a mixed radiation field consisting of $\gamma$-quanta, neutrons and ions (primarily protons and heavier fragments) during irradiation with a \SI{330.8}{\mega\electronvolt}/u $^{12}$C beam at the Marburg Ion-Beam Therapy Center (MIT) \cite{ref:mit}. 
The primary motivation for these measurements was to access the neutron (and $\gamma$-ray) doses to which patients might be exposed during FLASH irradiation \cite{ref:FLASH} protocol with carbon ions. Although performed as a side experiment alongside the main beam time allocation, the measurements provided a valuable opportunity to characterize both detector channels under realistic high-energy neutron conditions in the presence of a complex background of other secondary particles. 

\begin{figure}[!h]
    \includegraphics[width=0.48\textwidth]{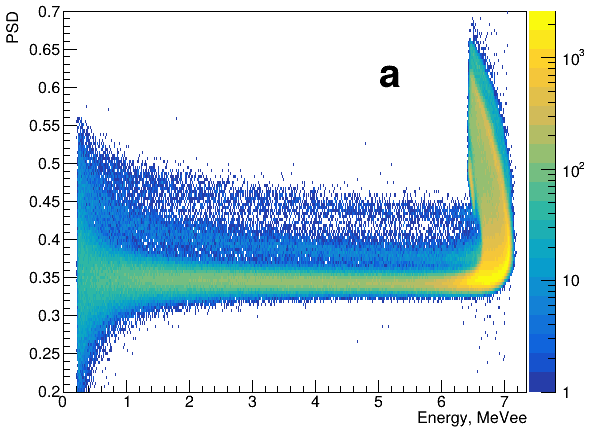}
    \includegraphics[width=0.47\textwidth]{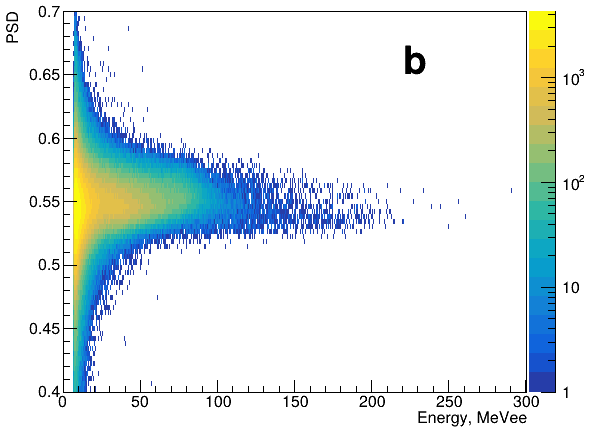}
\caption{\label{fig:mixed} 
Scatter-plot of PSD values versus deposited energy in $MeV_{ee}$ of the measured mixed field of $\gamma$-quanta, neutrons and ions obtained by low (a) and high (b) energy range channels. 
}
\end{figure}

The purpose of this measurement is to validate the dual-channel
architecture under high-energy and mixed particle field conditions that cannot be accessed with laboratory neutron sources. The resulting PSD versus energy histograms are presented in Figure~\ref{fig:mixed}.

The low-energy channel (Figure~\ref{fig:mixed}a) shows three distinct
bands whose qualitative features are consistent with the expected PSD
signatures for each particle class in EJ-276-type
scintillators~\cite{ref:ej276,ref:fast_neutron_psd}: the lowest band is
consistent with $\gamma$-ray responses, the intermediate band with proton
recoils from elastic (n,p) scattering, and the upper band with heavier
ions from beam fragmentation. This channel saturates above approximately
7\,MeV$_{ee}$, which is in line with its design range for laboratory
neutron sources and confirms that a second channel is essential for this
measurement environment.

The high dynamic range channel (Figure~\ref{fig:mixed}b) captures the
full energy spectrum of the secondary particle field without saturation,
covering the range well beyond 100\,MeV$_{ee}$. Even without dedicated
optimization for this specific radiation environment, the channel
successfully resolves distinct PSD populations, demonstrating that the
dual-channel readout concept functions as intended in a complex mixed
field. The differences in apparent energy limits and PSD band positions
relative to the laboratory source measurements are attributed to
variations in detector conditions between campaigns, including periodic
optical recoupling of the SiPMs.

A quantitative analysis of the secondary particle spectrum, including
particle identification supported by Monte Carlo simulation, is beyond
the scope of the present work and is planned for a future dedicated
beam-time campaign at MIT.

\section{Conclusion}
\label{Sec:Conclusion}

\begin{figure}[h]
\centerline{\includegraphics[width=0.5\textwidth]{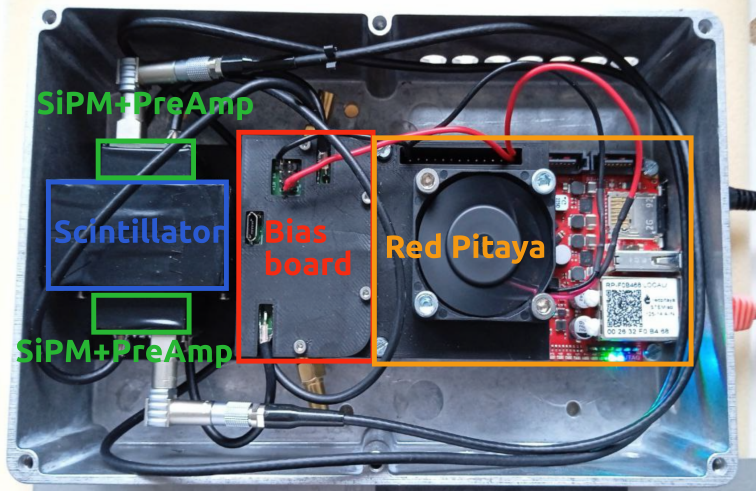}}

\caption{\label{fig:finaldetector} 
Compact solution of the final detector setup.
}
\end{figure}

In this work, we have demonstrated a compact, scalable, and
low-cost neutron detection system based on the EJ-276D plastic
scintillator and the StemLab Red Pitaya 125-14 platform
(Figure~\ref{fig:finaldetector}). The dual-channel SiPM readout with a
custom analog front-end based on the BGA61x MMIC family extends the
dynamic range from low-energy gamma calibration up to high-energy recoil
proton and heavy-ion signals.

The system achieves high-performance pulse shape discrimination (PSD), yielding a Figure of Merit (FoM) of 1.6 in the 2.75–3.0 $MeV_{ee}$ energy range. This performance is comparable to established liquid and plastic scintillation setups. Our solution is characterized by its extreme portability, requiring only a single +5 V power source and occupying less than half an A4 footprint, while maintaining a low power consumption of approximately 7.5 W.

While the current version of the prototype is optimized for event rates up to $5000\,\text{s}^{-1}$, it offers a robust solution for most environmental and laboratory radiation monitoring applications. 

The system is built entirely from commercially available, off-the-shelf
components: the Red Pitaya StemLab 125-14 is a general-purpose FPGA/DAQ
platform available from multiple distributors, the Hamamatsu SiPMs are standard distributor-stocked devices, and the custom PCBs can be produced by any PCB manufacturer. The complete detector assembly excluding the scintillator are therefore relatively inexpensive. This compares favorably with conventional fast neutron detector systems, which typically require specialized high-voltage NIM or VME modules and dedicated digitizer hardware---components that individually cost more than a thousand \texteuro\, per channel.

Future work will focus on further optimizing the FPGA firmware for higher event rates and exploring the scalability of this architecture for multi-detector arrays with different SiPM configurations using the four-input channel version of the Red Pitaya. Obtained results demonstrate that the Red Pitaya-based DAQ provides a versatile and cost-effective alternative for high-performance radiation detection without compromising on pulse discrimination quality.

The design files and firmware will be available upon request for the community to support the open-source development of compact instrumentation.

\section*{Acknowledgements}

The authors thank Dr. A. Abdulkhader and Dr. D. Krambrich for providing access to the AmBe source at the Radiation Center (JLU, Giessen).

\bibliography{bibliography}

\end{document}

%% file: preamp1.tex
\begin{tikzpicture}
	\draw (6, 7.312) to[cute inductor, /tikz/circuitikz/bipoles/length=1.12cm] (6, 6.312);
	\draw (3, 6.5) to[empty diode] (3, 7);
	\node[american buffer port] at (5.3, 6){};
	\node[ground] at (3, 4.578){};
	\draw (3, 5.828) to[american resistor, /tikz/circuitikz/bipoles/length=1.12cm] (3, 4.578);
	\draw (3, 6.5) -| (3, 6);
	\draw (6, 7.508) to[american resistor, /tikz/circuitikz/bipoles/length=1.12cm] (6, 8.747);
	\node[circ] at (3, 6){};
	\node[circ] at (6, 6){};
	\draw (3.01, 9.254) to[american resistor, /tikz/circuitikz/bipoles/length=1.12cm] (3.01, 10.494);
	\draw (3, 5.828) -- (3, 6);
	\draw (6, 9.267) to[american resistor, mirror, /tikz/circuitikz/bipoles/length=1.12cm] (6, 10.5);
	\draw (6, 9.25) -| (6, 8.75);
	\draw (7, 9) to[capacitor, /tikz/circuitikz/bipoles/length=0.980cm] (6.5, 9);
	\node[circ] at (6, 9){};
	\node[ground] at (7.501, 9){};
	\draw (6, 7.508) -- (6, 7.312);
	\draw (7, 6) to[capacitor, /tikz/circuitikz/bipoles/length=0.980cm] (6.5, 6);
	\draw (4.262, 6) to[capacitor, /tikz/circuitikz/bipoles/length=0.980cm] (3.762, 6);
	\draw (4.01, 9) to[capacitor, /tikz/circuitikz/bipoles/length=0.980cm] (3.51, 9);
	\node[ground] at (4.512, 9){};
	\node[shape=rectangle, minimum width=-0.035cm, minimum height=-0.035cm] at (6, 5){};
	\node[shape=rectangle, minimum width=1.676cm, minimum height=0.465cm] at (5.144, 5){} node[anchor=center, align=center, text width=1.218cm, inner sep=7pt] at (5.144, 5){\textcolor{rgb,255:red,0;green,0;blue,0}{BGA614}};
	\node[msport, xscale=-1.45, yscale=-1.45] at (5.693, 10.5){};
	\node[msport, xscale=-1.45, yscale=-1.45] at (2.48, 10.5){};
	\node[msport, xscale=-1.45, yscale=-1.45] at (8.476, 6){};
	\node[shape=rectangle, minimum width=0.754cm, minimum height=0.465cm] at (1.935, 10.5){} node[anchor=center, align=center, text width=0.296cm, inner sep=7pt] at (1.645, 10.5){\textcolor{rgb,255:red,0;green,0;blue,0}{$V_{Bias}$}};
	\node[shape=rectangle, minimum width=0.8cm, minimum height=0.465cm] at (5.082, 10.5){} node[anchor=center, align=center, text width=0.342cm, inner sep=7pt] at (4.982, 10.5){\textcolor{rgb,255:red,0;green,0;blue,0}{$V_{CC}$}};
	\node[shape=rectangle, minimum width=0.754cm, minimum height=0.465cm] at (7.921, 6.002){} node[anchor=center, align=center, text width=0.296cm, inner sep=7pt] at (7.721, 6.002){\textcolor{rgb,255:red,0;green,0;blue,0}{Out}};
	\node[shape=rectangle, minimum width=2.616cm, minimum height=0.465cm] at (1.334, 6.749){} node[anchor=center, align=center, text width=2.158cm, inner sep=7pt] at (1.834, 6.749){\textcolor{rgb,255:red,0;green,0;blue,0}{S14160-6050HS}};
	\node[shape=rectangle, minimum width=1.541cm, minimum height=0.465cm] at (6.83, 9.971){} node[anchor=west, align=left, text width=1.082cm, inner sep=7pt] at (6.042, 9.971){\textcolor{rgb,255:red,0;green,0;blue,0}{10 $\Omega$}};
	\node[shape=rectangle, minimum width=1.336cm, minimum height=0.465cm] at (5.303, 8){} node[anchor=west, align=left, text width=0.877cm, inner sep=7pt] at (4.618, 8){\textcolor{rgb,255:red,0;green,0;blue,0}{51 $\Omega$}};
	\node[shape=rectangle, minimum width=1.688cm, minimum height=0.465cm] at (2.403, 5.129){} node[anchor=west, align=left, text width=1.23cm, inner sep=7pt] at (1.541, 5.129){\textcolor{rgb,255:red,0;green,0;blue,0}{10 $\Omega$}};
	\node[shape=rectangle, minimum width=1.541cm, minimum height=0.465cm] at (6.896, 8.41){} node[anchor=west, align=left, text width=1.082cm, inner sep=7pt] at (6.108, 8.41){\textcolor{rgb,255:red,0;green,0;blue,0}{1 $\mu\mathrm{F}$}};
	\node[shape=rectangle, minimum width=1.541cm, minimum height=0.465cm] at (3.893, 8.41){} node[anchor=west, align=left, text width=1.082cm, inner sep=7pt] at (3.105, 8.41){\textcolor{rgb,255:red,0;green,0;blue,0}{1 $\mu\mathrm{F}$}};
	\node[shape=rectangle, minimum width=1.541cm, minimum height=0.465cm] at (6.817, 6.716){} node[anchor=west, align=left, text width=1.082cm, inner sep=7pt] at (6.029, 6.716){\textcolor{rgb,255:red,0;green,0;blue,0}{1 $\mu\mathrm{H}$}};
	\draw (3.01, 10.494) |- (2.48, 10.5);
	\draw (3.01, 9.254) |- (3, 9);
	\draw (3.51, 9) |- (3, 9);
	\node[circ] at (3.01, 9){};
	\draw (4.512, 9) |- (4.01, 9);
	\draw (3.762, 6) -- (3, 6);
	\draw (4.262, 6) -- (4.6, 6);
	\draw (6, 6) -- (6.5, 6);
	\draw (7, 6) -- (7.461, 6);
	\draw (7.501, 9) |- (7, 9);
	\draw (6.5, 9) -- (6, 9);
	\draw (6, 10.5) |- (5.693, 10.5);
	\draw (6, 9.267) -- (6, 9);
	\node[shape=rectangle, minimum width=1.541cm, minimum height=0.465cm] at (4.144, 5.42){} node[anchor=west, align=left, text width=1.082cm, inner sep=7pt] at (3.356, 5.42){\textcolor{rgb,255:red,0;green,0;blue,0}{1 $\mu\mathrm{F}$}};
	\node[shape=rectangle, minimum width=1.541cm, minimum height=0.465cm] at (6.896, 5.433){} node[anchor=west, align=left, text width=1.082cm, inner sep=7pt] at (6.108, 5.433){\textcolor{rgb,255:red,0;green,0;blue,0}{1 $\mu\mathrm{F}$}};
	\node[shape=rectangle, minimum width=1.689cm, minimum height=0.465cm] at (3.914, 9.958){} node[anchor=west, align=left, text width=1.23cm, inner sep=7pt] at (3.052, 9.958){\textcolor{rgb,255:red,0;green,0;blue,0}{10 $\mathrm{k}\Omega$}};
	\draw (3.01, 9) |- (3, 7);
	\draw (6, 6.312) -- (6, 6);
\end{tikzpicture}

%% file: preamp2.tex
\begin{tikzpicture}
	\draw (6, 7.484) to[cute inductor, /tikz/circuitikz/bipoles/length=1.12cm] (6, 6.484);
	\draw (3, 6.5) to[empty diode] (3, 7);
	\node[american buffer port] at (5.3, 6){};
	\node[ground, xscale=0.9, yscale=0.9] at (3, 4.274){};
	\draw (3, 5.828) to[american resistor, /tikz/circuitikz/bipoles/length=1.12cm] (3, 4.578);
	\draw (3, 6.5) -| (3, 6);
	\draw (6, 7.508) to[american resistor, /tikz/circuitikz/bipoles/length=1.12cm] (6, 8.747);
	\node[circ] at (3, 6){};
	\node[circ] at (6, 6){};
	\draw (3.01, 9.254) to[american resistor, /tikz/circuitikz/bipoles/length=1.12cm] (3.01, 10.494);
	\draw (3, 5.828) -- (3, 6);
	\draw (6, 9.267) to[american resistor, mirror, /tikz/circuitikz/bipoles/length=1.12cm] (6, 10.5);
	\draw (6, 9.25) -| (6, 8.75);
	\draw (7, 9) to[capacitor, /tikz/circuitikz/bipoles/length=0.980cm] (6.5, 9);
	\node[circ] at (6, 9){};
	\node[ground] at (7.501, 9){};
	\draw (7, 6) to[capacitor, /tikz/circuitikz/bipoles/length=0.980cm] (6.5, 6);
	\draw (4.262, 6) to[capacitor, /tikz/circuitikz/bipoles/length=0.980cm] (3.762, 6);
	\draw (4.01, 9) to[capacitor, /tikz/circuitikz/bipoles/length=0.980cm] (3.51, 9);
	\node[ground] at (4.512, 9){};
	\node[shape=rectangle, minimum width=-0.035cm, minimum height=-0.035cm] at (6, 5){};
	\node[shape=rectangle, minimum width=1.676cm, minimum height=0.465cm] at (5, 5){} node[anchor=center, align=center, text width=1.218cm, inner sep=7pt] at (5, 5){\textcolor{rgb,255:red,0;green,0;blue,0}{BGA616}};
	\node[msport, xscale=-1.45, yscale=-1.45] at (5.693, 10.5){};
	\node[msport, xscale=-1.45, yscale=-1.45] at (2.48, 10.5){};
	\node[msport, xscale=-1.45, yscale=-1.45] at (8.476, 6){};
	\node[shape=rectangle, minimum width=0.754cm, minimum height=0.465cm] at (1.935, 10.5){} node[anchor=center, align=center, text width=0.296cm, inner sep=7pt] at (1.645, 10.5){\textcolor{rgb,255:red,0;green,0;blue,0}{$V_{Bias}$}};
	\node[shape=rectangle, minimum width=0.8cm, minimum height=0.465cm] at (5.082, 10.5){} node[anchor=center, align=center, text width=0.342cm, inner sep=7pt] at (4.982, 10.5){\textcolor{rgb,255:red,0;green,0;blue,0}{$V_{CC}$}};
	\node[shape=rectangle, minimum width=0.754cm, minimum height=0.465cm] at (7.921, 6.002){} node[anchor=center, align=center, text width=0.296cm, inner sep=7pt] at (7.821, 6.002){\textcolor{rgb,255:red,0;green,0;blue,0}{Out}};
	\node[shape=rectangle, minimum width=2.616cm, minimum height=0.465cm] at (1.334, 6.749){} node[anchor=center, align=center, text width=2.158cm, inner sep=7pt] at (1.834, 6.749){\textcolor{rgb,255:red,0;green,0;blue,0}{S14160-3010PS}};
	\node[shape=rectangle, minimum width=1.541cm, minimum height=0.465cm] at (6.83, 9.971){} node[anchor=west, align=left, text width=1.082cm, inner sep=7pt] at (6.042, 9.971){\textcolor{rgb,255:red,0;green,0;blue,0}{10 $\Omega$}};
	\node[shape=rectangle, minimum width=1.336cm, minimum height=0.465cm] at (5.303, 8){} node[anchor=west, align=left, text width=0.877cm, inner sep=7pt] at (4.618, 8){\textcolor{rgb,255:red,0;green,0;blue,0}{17 $\Omega$}};
	\node[shape=rectangle, minimum width=1.465cm, minimum height=0.465cm] at (2.25, 5.129){} node[anchor=east, align=right, text width=1.006cm, inner sep=7pt] at (3, 5.129){\textcolor{rgb,255:red,0;green,0;blue,0}{220 $\Omega$}};
	\node[shape=rectangle, minimum width=1.541cm, minimum height=0.465cm] at (6.896, 8.41){} node[anchor=west, align=left, text width=1.082cm, inner sep=7pt] at (6.008, 8.41){\textcolor{rgb,255:red,0;green,0;blue,0}{1 $\mu\mathrm{F}$}};
	\node[shape=rectangle, minimum width=1.541cm, minimum height=0.465cm] at (3.893, 8.41){} node[anchor=west, align=left, text width=1.082cm, inner sep=7pt] at (3.105, 8.41){\textcolor{rgb,255:red,0;green,0;blue,0}{1 $\mu\mathrm{F}$}};
	\node[shape=rectangle, minimum width=1.936cm, minimum height=0.465cm] at (5.38, 7){} node[anchor=west, align=left, text width=1.477cm, inner sep=7pt] at (4.394, 7){\textcolor{rgb,255:red,0;green,0;blue,0}{470 $\mathrm{nH}$}};
	\draw (3.01, 10.494) |- (2.48, 10.5);
	\draw (3.01, 9.254) |- (3, 9);
	\draw (3.51, 9) |- (3, 9);
	\node[circ] at (3.01, 9){};
	\draw (4.512, 9) |- (4.01, 9);
	\draw (3.762, 6) -- (3, 6);
	\draw (4.262, 6) -- (4.6, 6);
	\draw (6, 6) -- (6.5, 6);
	\draw (7, 6) -- (7.461, 6);
	\draw (7.501, 9) |- (7, 9);
	\draw (6.5, 9) -- (6, 9);
	\draw (6, 10.5) |- (5.693, 10.5);
	\draw (6, 9.267) -- (6, 9);
	\node[shape=rectangle, minimum width=1.541cm, minimum height=0.465cm] at (4.144, 5.42){} node[anchor=west, align=left, text width=1.082cm, inner sep=7pt] at (3.256, 5.42){\textcolor{rgb,255:red,0;green,0;blue,0}{1 $\mu\mathrm{F}$}};
	\node[shape=rectangle, minimum width=1.541cm, minimum height=0.465cm] at (6.962, 6.5){} node[anchor=west, align=left, text width=1.082cm, inner sep=7pt] at (6.174, 6.5){\textcolor{rgb,255:red,0;green,0;blue,0}{0.1 $\mu\mathrm{F}$}};
	\node[shape=rectangle, minimum width=1.689cm, minimum height=0.465cm] at (3.914, 9.958){} node[anchor=west, align=left, text width=1.23cm, inner sep=7pt] at (3.052, 9.958){\textcolor{rgb,255:red,0;green,0;blue,0}{10 $\mathrm{k}\Omega$}};
	\draw (3.01, 9) |- (3, 7);
	\draw (6, 6.312) -- (6, 6);
	\draw (6, 4.927) to[american resistor, /tikz/circuitikz/bipoles/length=0.840cm] (6, 5.816);
	\node[ground, xscale=0.9, yscale=0.9] at (5.99, 4.287){};
	\draw (6, 5.816) -- (6, 6);
	\draw (5.996, 4.848) to[capacitor, /tikz/circuitikz/bipoles/length=0.840cm] (5.996, 4.348);
	\draw (6, 4.927) |- (5.996, 4.848);
	\draw (5.996, 4.348) |- (5.99, 4.287);
	\draw (3, 4.578) -- (3, 4.274);
	\node[ground, xscale=0.9, yscale=0.9] at (1.399, 4.274){};
	\draw (2.907, 6) -- (1.407, 6) |- (1.405, 5.377);
	\draw (1.392, 4.877) -| (1.399, 4.274);
	\node[shape=rectangle, minimum width=1.541cm, minimum height=0.465cm] at (0.599, 5.102){} node[anchor=west, align=left, text width=1.082cm, inner sep=7pt] at (-0.189, 5.102){\textcolor{rgb,255:red,0;green,0;blue,0}{33 $\mathrm{pF}$}};
	\node[shape=rectangle, minimum width=1.872cm, minimum height=0.465cm] at (7.046, 4.579){} node[anchor=west, align=left, text width=1.414cm, inner sep=7pt] at (6.093, 4.579){\textcolor{rgb,255:red,0;green,0;blue,0}{100 $\mathrm{pF}$}};
	\node[shape=rectangle, minimum width=1.336cm, minimum height=0.465cm] at (6.711, 5.343){} node[anchor=west, align=left, text width=0.877cm, inner sep=7pt] at (6.025, 5.343){\textcolor{rgb,255:red,0;green,0;blue,0}{10 $\Omega$}};
	\draw (6, 6.484) -- (6, 6);
	\draw (6, 7.508) -- (6, 7.484);
	\draw (1.407, 5.377) to[capacitor, /tikz/circuitikz/bipoles/length=0.980cm] (1.399, 4.877);
\end{tikzpicture}